\documentclass[showpacs,preprintnumbers,amsmath,amssymb]{revtex4}

\def\ben{\begin{enumerate}} \def\een{\end{enumerate}}
\def\beq{\begin{equation}} \def\eeq{\end{equation}}
\def\bea{\begin{eqnarray}} \def\eea{\end{eqnarray}}
\def\beann{\begin{eqnarray*}} \def\eeann{\end{eqnarray*}}
\def\beasn{\begin{sneqnarray}} \def\eeasn{\end{sneqnarray}}

\begin{document}


\title{Gauge Fixing and Observables in General Relativity}

\author{D.\ C.\ Salisbury
\footnote[2]{Balfest Conference Proceedings,
Vietri sul Mare, 26-31 May 2003}}
\address{Department of Physics,
Austin College, Sherman, Texas 75090-4440, USA,
dsalisbury@austincollege.edu}

\date{19 October 2003}

\begin{abstract}
The conventional group of four-dimensional diffeomorphisms is not 
realizeable as a canonical transformation group in phase space. 
Yet there is a larger field-dependent symmetry transformation group 
which does faithfully reproduce 4-D diffeomorphism symmetries. Some 
properties of this group were first explored by Bergmann and Komar. 
More recently the group has been analyzed from the perspective of 
projectability under the Legendre map. Time translation is not 
a realizeable symmetry, and is therefore distinct from 
diffeomorphism-induced symmetries. This issue is explored 
further in this paper. It is shown that time is not "frozen". 
Indeed, time-like diffeomorphism invariants must be time-dependent. 
Intrinsic coordinates of the type proposed by Bergmann 
and Komar are used to construct invariants. Lapse and shift variables are 
retained as canonical variables in this approach, and therefore 
will be subject to quantum fluctuations in an eventual quantum 
theory. Concepts and constructions are illustrated using the 
relativistic classical and quantum free particle. In this example 
concrete time-dependent invariants are displayed and fluctuation 
in proper time is manifest. 

\end{abstract}

\pacs{4.20.Fy, 4.60.Ds. \hfill gr-qc/0310095}

\maketitle

\section{Homage and Introduction}	

What better way to celebrate the birthday of our dear friend and
teacher Balachandran than to continue to reflect and debate together, 
with the same intellectural rigor, impassioned commitment and
irreverent playfulness most of us recall from the decades-old
tradition of Physics Building Room 316. And what a joy it has been to
gather together generations of physicists who were formed by this
process. Thank you Bal for your inspirational example. Alas, though
your explorations in quantum geometry have long stimulated my own
thinking about the nature of time, I'm afraid I never did learn
master your skill in using time wisely!

My collaborators J.M.Pons, L. C. Shepley, and I have recently
elucidated the nature of four-dimensional diffeomorphism symmetries
in canonical phase space formulations of general 
relativity\cite{pss:1997pr,pss:2000jmp,pss:2000grg,pss:2000pr}. It turns
out that a global constant translation in time (time evolution) is
not realizeable as a symmetry transformation. Thus symmetries and
time evolution are mathematically distinct. The incorrect
indentification of time evolution as a symmetry has led to the so
called ``problem of time'', the assertion that diffeomorphism
invariants must be independent of time. Various routes to this
mistaken
conclusion are analyzed by J. M. Pons and myself in a forthcoming paper.
\cite{ps03} I will review here the general framework, with special
emphasis on the retention of lapse and shift variables as canonical variables.
They play an essential role in the diffeomorphiem-induced symmetry
transformation group, and in addition there is good reason for
promoting them to quantum variables subject to fluctuation. This
property will be displayed explicitly in a free relativistic particle
example. Diffeomorphism invariants will be constructed through the
use of intrinsic coordinates, coordinates whose values are fixed
by the values of physical fields. These invariants in general
relativity are necessarily time dependent.

\section{Legendre Projectability of Diffeomorphism Symmetries}

All generally covariant theories have the property that
variations of physical variables generated by diffeomorphisms which
alter the time are not projectable under the Legendre transformation 
to phase space. I will give the generic explanation and then
illustrate with the relativistic free particle. Suppose we have a
system described by a Lagrangian $L(q, \dot q)$, with configuration
variables $q^{i}$ and velocities $\dot q^{i}$. Then due to the
general covariance of the model the Legendre matrix is singular:
$\det{\frac{\partial^{2} L}{\partial \dot q^{i} \partial \dot q^{j}}} = 0$. 
Thus functions $f(q, \dot q)$ which vary along the null directions of
this matrix are not projectable under the Legendre map 
$p_{i} =\frac{\partial L}{\partial \dot q^{i}}$, as I now show. Let
$\gamma^{i}$ represent the components of a null vector, i.e., 
$\frac{\partial^{2} L}{\partial \dot q^{i} \partial \dot q^{j}}\gamma^{i} =0$. 
More precisely I mean by ``projectable'' that $f$ is the pullback of 
a function $F(q,p)$ in phase space. If this is the case it follows
that
\beq
\gamma^{i} \frac{\partial f(q, \dot q)}{\partial \dot q^{i}} =
\frac{\partial F(q, p(q, \dot q))}{\partial p_{k}} \gamma^{i} 
\frac{\partial^{2} L}{\partial \dot q^{k} \partial \dot q^{i}} = 0.
\eeq
and the assertion is proved. Consider as an example the relativisitc 
free particle described by the Lagrangian
$L = \frac{1}{N} \dot x^{2} - \frac{N}{2}$, where $x^{\mu}(t)$ is the
Minkowski spacetime position, dependent  on the arbitrary parameter
$t$. The variable $N$ is the lapse which determines the proper time
elapsed as a function of $t$: $d\tau = N(t) dt$. 
The resulting equations of motion are covariant under arbitrary
changes in the parameterization. Consequently, the Legendre matrix
possesses a null direction, given in this case by 
$\frac{\partial}{\partial \dot N}$, so projectable functions may not 
depend on $\dot N$.

In fact, the time derivatives of the lapse and shift functions are
absent in all generally covariant theories, including general
relativity. The lapse $N$ and shift $N^{a}$ appear in the $3+1$
decomposition of the spacetime metric
\beq
g_{\mu\nu}
    = \left(\begin{array}{cc}
	-N^{2}+g_{cd}N^{c}N^{d}  \quad    & g_{ac} N^c \\
	g_{bc} N^c  & g_{ab}
    \end{array}\right)\ .
	\label{themetric} 
\end{equation}
Let us calculate the variations of the metric (the Lie derivative) 
induced by infinitesimal
coordinate transformations of the form 
$x'^{\mu} = x^{\mu} -\epsilon^{\mu}(x)$, where the $\epsilon^{\mu}$
are arbitrary infinitesmal functions, 
\beq
\delta g_{\mu \nu} = g_{\mu \nu , \alpha} \epsilon^{\alpha} + g_{\alpha \nu
}\epsilon^{\alpha}_{,\mu} + g_{\mu \alpha }\epsilon^{\alpha}_{,\nu}.
\eeq
We note that variations of the lapse and shift do depend on these time
derivatives. Therefore all these variations corresponding to non-vanishing
$\epsilon^{0}$ are not projectable.
For example, we find in our free particle example that under an arbitrary
infinitesmal reparameterization $t' = t - \epsilon(t)$ the variation 
of $N$ is $\delta N = \dot N \epsilon + N \dot \epsilon $.

Nevertheless there does exist a gauge symmetry group which reproduces
on any given solutions of the equations of motion all symmetry
transformations engendered by the conventional diffeomorphism group. 
This trick is accomplished by broadening the group to include a
compulsory dependence on the lapse and shift. It is straightforward
to prove that the 
unique form of $\epsilon$ which produces projectable variations of the lapse and
shift is \cite{pss:1997pr},
\beq
\epsilon^{\mu}(x, N, N^{a}) = \delta^{\mu}_{a} \xi^{a}(x) + n^{\mu}(x)
\xi^{0}(x),
\eeq
where $n^{\mu}$ is the orthogonal vector to the time foliation of
spacetime and is given by
\beq
n^{\mu} = (N^{-1}, - N^{-1} N^{a}).
\eeq
The $\xi^{\mu}$ are arbitrary infinitesmal functions of the
coordinates.
This familiar decomposition was first introduced by Dirac, though the
group theoretical explanation was not known until decades later. In
fact this form is not quite correct since, as was first pointed out
by Bergmann and Komar \cite{berg-kom72}, the group multiplication rule in general
relativity produces an unavoidable non-local dependence of $\xi^{0}$ 
on the spatial metric . Returning to our example we note 
that under 
$t'= t - N^{-1} \xi(t)$ we find
\beq
\delta N = \dot N N^{-1} \xi + N \frac{d}{dt}\left(N^{-1} \xi \right)
= \dot \xi.
\eeq

\section{Symmetry Generators and the Hamiltonian}

The general structure of the generators of diffeomorphism-induced
symmetries, in which the lapse and shift are retained as canonical
variables, is
\beq
G[\xi] = \int d^{3}x \, \left( \dot \xi^{\mu} P_{\mu} +
\xi^{\mu} \left( {\cal H}_{\mu} + \int \int d^{3}y \, d^{3}z \,
C^{\beta}_{\mu \alpha}(x,y,z) N^{\alpha}(y) P_{\beta})\right) \right).
\eeq
In this expression the $P_{\mu}$ are the momenta conjugate to the
lapse and shift. They are constrained to vanish. Preservation of
these constraints in time leads to the secondary constraints 
${\cal H}_{\mu} \approx 0$. These constraints form a closed equal-time Poisson
Bracket algebra with structure functions $C^{\beta}_{\mu \alpha}$:
\beq
\{ {\cal H}_{\mu}(x^{0},\vec x) , {\cal H}_{\nu}( x^{0},\vec y)
\}_{PB} = \int d^{3}z \, C^{\beta}_{\mu \nu}(x,y,z) {\cal H}_{\beta}.
\eeq
Some $C$'s depend on the spatially differentiated three-metric and
this is the origen of
the non-local group dependence on the three-metric noted above. The
generator in our free particle example is
\beq
G[\xi] = \dot \xi P + \xi \frac{1}{2} (p^{2} +1). \label{pargen}
\eeq
These generators 
are distinct from the Hamiltonian which is of the form
\beq
H = \int d^{3}x \, (N^{\mu} {\cal H}_{\mu} + \lambda^{\mu} P_{\mu}).
\eeq
In this expression $\lambda(x)$ are monotonically increasing
functions of $x^{0}$ but otherwise arbitrary. Note that in
contrast with the symmetry generator $G[\xi]$ the
secondary constraints in $H$ are multiplied by the canonical
variables $N^{\mu}$, and not by arbitrary functions. Note also that
a time-dependent choice of $\lambda$ implies a time-dependent
Hamiltonian. The Hamiltonian 
for the free particle is $H = \frac{N}{2}(p^{2} + 1) + \lambda P$.

\section{Finite Time Evolution and Symmetry Transformations}

The finite time evolution operator which evolves data from time zero 
to $t$ is
\bea
\hat U(t,0) &=& {\cal T} \exp \left( \int_{0}^{t} dt' \, \{\quad ,
H(t') \}_{PB} \right) \nonumber \\
&=& 1 + \int_{0}^{t} dt' \, \{\quad ,H(t') \}_{PB} \nonumber \\
&+&
\int_{0}^{t} dt_{1} \int_{0}^{t_{1}} dt_{2}\, 
\{ \{\quad ,H(t_{1})\}, H(t_{2} \}_{PB} + \dots
\eea
In the first line ${\cal T}$ denotes time ordering. It is straightforward
to compute the time evolution of variables in our free particle
model. We let the variables with no time argument denote their
initial values (and hence their initial phase space coordinate). In
addition to $p_{\mu}(t) = p_{\mu}$ we find
\beq
N(t) = \hat U(t,0) N = N + \int_{0}^{t} dt_{1} \, \lambda (t_{1}),
\label{nsol}
\eeq
and
\beq
x^{\mu}(t) = \hat U(t,0) x^{\mu} = x^{\mu} + \int_{0}^{t} dt_{1}
N(t_{1}) p^{\mu}. \label{xsol}
\eeq

The finite symmetry operator $\hat S(s)$ can now be applied to
solutions to create a one-parameter family of gauge-transformed
solutions associated with the finite group descriptors $\xi$. If
$\xi$ is chosen to be independent of the group parameter $s$ we have
\beq
\hat S(s) = \exp \left(s \{ \quad ,G[\xi] \}_{PB} \right).
\eeq
( An $s$ dependence in $\xi$ would imply an $s$ ordering in
this expression. It turns out that such an $s$ dependence can be
exploited to simplify the resulting general coordinate transformation
\cite{ps03b}.) In our free particle example 
$\hat S(s) = \frac{\xi}{2} (p^{2} + 1) + \dot \xi P$, and it
generates the following one-parameter family of gauge transformed
solutions:
\beq
N_{s}(t) := \hat S(s) N(t) = N(t) + s \dot \xi(t), \label{nssol}
\eeq
and
\beq
x^{\mu}_{s}(t) := \hat S(s) x^{\mu}(t) = x^{\mu}(t) + s \xi(t) p^{\mu}.
\label{xssol}
\eeq

It is evident in this example that the symmetry transformation
effectively alters the original choice of $\lambda$ in the
Hamiltonian. In general we note we now have at our disposal a
symmetry transformation which arbitrarily alters the original choice 
of time foliation. That is, in spite of having been forced to choose a 
foliation which appears to destroy the full four-dimensional
diffeomorphism symmetry, this symmetry is indeed present after all
and can in principle be implemented in canonically quantized models!

\section{Gauge Fixing and Intrinsic Coordinates}

We will construct invariants under the diffeomorphism-induced group
through the use of intrinsic coordinates. This is a program first
proposed by Einstein himself in reconciling himself with the concept 
of general covariance. The dilemma he faced and his resolution is
nicely reviewed in J. Stachel's discussion of the ``hole argument''
\cite{stachel89}. 
More recently Komar and Bergmann proposed the use of Weyl scalars in 
the definition of intrinsic coordinates. In addition they showed how 
these scalars can be expressed on solution trajectories in terms of
canonical phase space variables \cite{berg-kom60}.

Since by assumption the prescription for passage to intrinsic
coordinates is unique, all observers (independent of their initial
coordinate choice) will agree on the numerical values of all
geometric objects provided  their solutions lie on the same
diffeomorphism-induced equivalence class of solutions. The
operational procedure a user would implement would be to set
intrinsic coordinates equal to a prescribed set of functions of
scalars formed from the physical fields. Then a coordinate
transformation of physical variables would be undertaken from the original 
arbitrarily
chosen coordinate system to the intrinsic coordinate system.  

This procedure can equivalently be viewed as a gauge fixing. From
this point of view we could for instance, in our relativisitc free
particle example, set
the parameter time coordinate equal to a function of the
reparametrization scalar $x^{0}(t)$,
\beq
t - f^{-1}(x^{0}(t)) \approx 0,
\eeq
where $f$ is a monotonically increasing function.
Given an arbitrary initial parameterization this constraint will not 
generically be fulfilled. But because we can now implement a canonical symmetry
transformation we can move this solution along the gauge orbit to the
solution which does satisfy the gauge condition constraint. The
required finite one-parameter family descriptor will then depend on
the original solution. (We will  take the parameter value $s = 1$).
The resulting gauge transformed solution is by construction a
diffeomorphism-induced invariant, i.e., an observable; a variation of
the original coordinization (which does not satisfy the gauge
constraint) results in no change in the gauge-transformed solution.
We will now carry out this procedure in detail.

Using (\ref{xssol}) we set the gauge-transformed $x_{s=1}^{0}(t)$ equal
to $f(t)$
\beq
f(t) = x^{0}(t) + \xi[x](t) p^{0}.
\eeq
Thus we find that the required solution-dependent descriptor is
$\xi[x](t) = \frac{1}{p^{0}}(f(t)-x^{0}(t))$. Substituting this
descriptor into (\ref{xssol}) we get the invariants
\beq
x^{a}_{\xi[x]}(t) = x^{a}(t) + \frac{p^{a}}{p^{0}}\left( f(t) - x^{0}(t)
\right) = x^{a} + \frac{p^{a}}{p^{0}}\left( f(t) - x^{0}
\right),
\eeq
and
\beq
N_{\xi[x]}(t) = N(t) + \dot \xi[x](t) =
\frac{1}{p^{0}}\frac{df(t)}{dt}. \label{ninv}
\eeq
These functions, in addition to $x^{0}(t) = f(t)$
are invariant under diffeomorphism-induced symmetry
transformation as we now show explicitly. 

Before applying the symmetry generator (\ref{pargen}) 
we need to observe that the time evolved constraint variable 
$P(t) = P- \frac{1}{2} (p^{2} + 1) t$. Then
the only non-vanishing symmetry
variations generated by $G[\eta]$ for infinitesmal $\eta$ are
\beq
\delta x^{\mu} = \{ x^{\mu}, \dot \eta(t) P(t) + \eta(t)  \frac{1}{2} (p^{2} +1) 
\}_{PB} = (\eta(t) -t \dot \eta (t)) p^{\mu},
\eeq
and $\delta N = \dot \eta $. Therefore since $x^{0}_{\xi[x]}(t) = f(t)$ and 
$N_{\xi[x]}(t) = \frac{1}{p^{0}}\frac{df(t)}{dt}$ do not depend on
these canonical variables they are trivially invariant. In addition,
\beq
\delta x^{a}_{\xi[x]}(t) = \delta x^{a}  - \frac{p^{a}}{p^{0}} \delta
x^{a} = 0.
\eeq

We have here an explicit construction of invariants which depend on
time. Invariants in general relativity will share this characteristic.
In fact, we now reinterpret a theorem due to Torre \cite{torre93} as a proof that
no {\it time-independent} invariants exist in general relativity since 
the theory contains no additional symmetry beyond general covariance.

\section{Implications for Quantum Gravity}

The program I have described can in principle be applied in general
relativity using Weyl scalars to fix intrinsic coordinates. There is 
of course a significant practical problem which must be addressed
even at the classical level: it will be necessary to find a general
temporally 
monotonically increasing function of Weyl scalars. It is probable
that such functions can only be found in coordinate patches which
must then be pieced together. Or one may perhaps be forced to abandon
the attempt to define time purely gravitationally and one may need to
resort to the use of material fields.

On the other hand it should now be clear that quantum time evolution 
can be given a sensible meaning. We are led to consider an improved
Wheeler-DeWitt formalism which takes this evolution into account. A
Hamilton-Jacobi approach which retains the lapse and shift is also an
attractive candidate for making the transition to quantum mechanics. 
Regarding the use of lapse and shift, there are two paths we might
contemplate. We could perform a group average over the
diffeomorphism-induced symmetry group. Or we could pursue the gauge
fixing formalism outlined in this paper and solve for the metric
variables, including lapse and shift, in terms of the true degrees of
freedom of the gravitational field.

Before embarking on the quantum implementation of this latter program in our
free particle model it will be instructive to review the advantages
the retention of lapse and shift will yield in quantum gravity. On 
the one hand it will in principle be possible to convert from one
choice of intrinsic coordinates to another. Of course, to do so we
must find a suitable choice of factor ordering which preserves the
gauge symmetry algebra. But perhaps more importantly, in addition to 
our ability to alter the quantum time foliation, we will introduce
quantum fluctuation in the full spacetime metric. Current connection 
approaches to quantum gravity have found a discretization of
spacelike 2-surfaces and volumes \cite{rovelli-smolin95}. 
A quantum theory which truly
reflects an underlying four dimensional diffeomorphism symmetry (if
only with respect to a small deformation of the time foliation) must
possess a timelike discretization. Operators representing timelike
surfaces can be constructed in the framework of the connection
approach, but they require additional physical geometric objects,
namely the lapse and shift in addition to the temporal component of
the connection. The classical symmetry analysis has been extended to 
the connection formalism (with arbitrary Immirzi parameter) 
\cite{pss:2000pr,ps02}, and work
on canonical quantization is in progress.

Let us briefly return to the free particle. We can carry out a
conventional quantization in which the one-particle Hilbert space is 
spanned by a spatial  momentum basis $|\vec p>$. We make 
the intrinsic time choice $x^{0}(t) = f(t)$. This means in practice
that observers have rate-adjusted their clocks in this manner. Now we
note from (\ref{ninv}) that the quantum lapse operator is given by
\beq
\hat N (t) = \frac{1}{\sqrt{\hat {\vec p}^{2} + 1}
}\frac{df(t)}{dt},
\eeq
so the elapsed proper time 
\beq
 \Delta \hat \tau = \int_{t_{1}}^{t_{2}} dt \hat N(t) =
\left( f(t_{2}) - f(t_{1}) \right) \frac{1}{\sqrt{\hat {\vec p}^{2} +
1}},
\eeq
is subject to quantum fluctuation!

\section{Conclusions}

We have observed that canonical classical general relativity is
covariant under symmetry transformations which are induced by the
full four-dimensional diffeomorphism symmetry group.
Misunderstandings about the nature of this group have led some to the
mistaken conclusion that diffeomorphism invariants must be constant
in time. In fact, quite the opposite is true. In general relativity
invariants exist and cannot be constant in time. Such invariants can 
be constructed through a choice of intrinsic coordinates and we have 
displayed an example of time-dependent invariants for the relativistic free
particle. Similar misconceptions have led to the mistaken conclusion 
that a choice of time foliation of spacetime leaves only the spatial 
diffeomorphism group as the remaining symmetry group. This is not the
case, and one can in fact implement symmetry transformations which 
arbitrarily 
alter the foliation. Finally, there is good physical rationale for
retaining the lapse and shift as classical and quantum variables.
They must be retained to exploit the full four-dimensional symmetry, 
and in the quantum model they become subject to a desireable quantum fluctuation.

\section*{Acknowledgments}

Thank you Bal for having shown me many years ago, amongst many other 
marvels, how to construct canonical invariants through the imposition of gauge
conditions!


\end{document}